\documentclass{aastex}          
\usepackage{spr-astr-addons}    

\def\lsim{\mathrel{\rlap{\lower 3pt \hbox{$\sim$}} \raise 2.0pt \hbox{$<$}}}
\def\gsim{\mathrel{\rlap{\lower 3pt \hbox{$\sim$}} \raise 2.0pt \hbox{$>$}}}
\def\pd{{\rm pd}}
\def\dv{\Delta{\rm V}}
\def\ad{\Delta\theta}

\begin{document}

\title{Reclassification of the nearest quasar pair candidate:\\ 
SDSS~J15244+3032 -- RXS~J15244+3032}

\shorttitle{Reclassification of the nearest quasar pair}
\shortauthors{Farina et al.}

\author{E.~P.~Farina}
\affil{Universit\`a degli Studi dell'Insubria ---  via Valleggio 11, I-22100 Como, Italy}
\affil{INFN Milano--Bicocca --- Universit\`a degli Studi di Milano--Bicocca, Piazza della Scienza 3, I-20126 Milano, Italy}
\email{emanuele.paolo.farina@gmail.com}
\and
\author{R.~Falomo}
\affil{INAF --- Osservatorio Astronomico di Padova, Vicolo dell'Osservatorio 5, I-35122 Padova, Italy}
\and
\author{A.~Treves\altaffilmark{1}}
\affil{Universit\`a degli Studi dell'Insubria ---  via Valleggio 11, I-22100 Como, Italy}
\affil{INFN Milano--Bicocca --- Universit\`a degli Studi di Milano--Bicocca, Piazza della Scienza 3, I-20126 Milano, Italy}
\and
\author{R.~Decarli}
\affil{Max--Planck--Institut f{\"u}r Astronomie --- K{\"o}nigstuhl 17, D-69117 Heidelberg, Germany}
\and
\author{J.~Kotilainen}
\affil{Finnish Centre for Astronomy with ESO (FINCA) --- University of Turku, V{\"a}is{\"a}l{\"a}ntie 20, FI-21500 Piikki{\"o}, Finland}
\and
\author{R.~Scarpa}
\affil{Instituto de astrof{\`i}sica de Canarias --- c/via Lactea s/n, E-38205 San Cristobal de la Laguna, Spain}

\altaffiltext{1}{Associated to INAF}

\begin{abstract}
We present optical spectroscopy of the nearest quasar pair listed 
in the 13$^{\rm th}$ edition of the V{\'e}ron-Cetty \& V{\'e}ron 
catalogue, i.e. the two quasars SDSS~J15244+3032 and RXS~J15244+3032 
(redshift $z\approx0.27$, 
angular separation $\Delta\theta\approx7\arcsec$,
and line--of--sight velocity difference $\dv\approx1900$\,km/s).
This system would be an optimal candidate to investigate the mutual 
interaction of the host galaxies with ground based optical imaging 
and spectroscopy. However, new optical data demonstrate that 
RXS~J15244+3032 is indeed a star of spectral type~G.\\

This paper includes data gathered with the Asiago 1.82\,m telescope 
(Cima Ekar Observatory, Asiago, Italy).
\end{abstract}


\section{Introduction}

In the last years, increasing attention has been given to the search for quasar 
pairs (i.e., two quasars close in the sky and with almost the same redshift) 
in order to assess the role of galaxy interactions in quasar ignition 
\citep[e.g.,][]{Hennawi2006, Hennawi2010, Myers2007a, Myers2007b, Myers2008, 
Foreman2009, Shen2010, Decarli2010, Farina2011, Kayo2012, Richardson2012}.
However, the discovery of these systems is challenging. Even the large 
spectroscopic quasar catalogue of \citet{Schneider2010} which holds more than 
$\sim100000$ objects from the 7th data release of the Sloan Digital Sky Survey 
\citep[SDSS,][]{Abazajian2009}, contains 22 quasars with angular separation 
$\ad<15\arcsec$, and only 12 with $\ad<10\arcsec$. This arises mainly as an
effect due to the so--called fiber collision limit: the finite size of the 
fiber plugs prevents the collection of spectra of more than one object with 
separation below $55\arcsec$ within the same SDSS plate \citep[][]{Blanton2003}.

In order to increase the number of known small separation pairs, we have 
scrutinised the 13$^{\rm th}$ edition of the V{\'e}ron-Cetty \& V{\'e}ron 
quasar catalogue \citep[][hereafter, VCV10]{Veron2010}, where the effects of 
the fiber collision limits should be mitigated by the inhomogeneous origin of 
the listed sources. In this paper we present compelling evidences that one 
of the discovered systems: SDSS~J15244+3032 and RXS~J15244+3032 with 
$\ad=7\farcs6$ is instead a quasar--star pair. 

Throughout this paper we consider a concordance cosmology with 
H$_0=70$\,km/s/Mpc, $\Omega_{\rm m} = 0.3$, and $\Omega_\Lambda=0.7$.

\section{J1524QS}\label{sec:J1524QS}

With the aim of discovering new close physical pairs of quasars we 
investigated the VCV10 catalogue which contains more than $130000$ 
quasars with known redshifts.
We found 58 pairs of quasars with projected separation $\pd\leq50$~kpc 
(at the redshift of the nearer quasar), 28 of which are physical pairs 
\citep[i.e. have line--of--sight velocity difference $\dv\leq2000$\,km/s,
see][]{Hennawi2006} and thus are well suited to the study of interacting 
systems. 
In fact the explored projected scales ($\pd<50$~kpc) are those where the 
quasar activity is expected to be triggered by dissipative interaction 
events, and the probability that they are due to chance superpositions 
is rather low \citep{Hennawi2006, Myers2007a, Myers2007b, Foreman2009, 
Farina2011, Kayo2012}. Of particular interest are the lower 
redshift systems, for which ground based data allow direct investigation 
of the effects of mutual interaction and of the galactic environment on 
triggering the quasar activity \citep[e.g.,][]{Mortlock1999, Green2010, 
Green2011}.

\begin{figure}[tb]
  \centering
  \includegraphics[width=0.99\columnwidth]{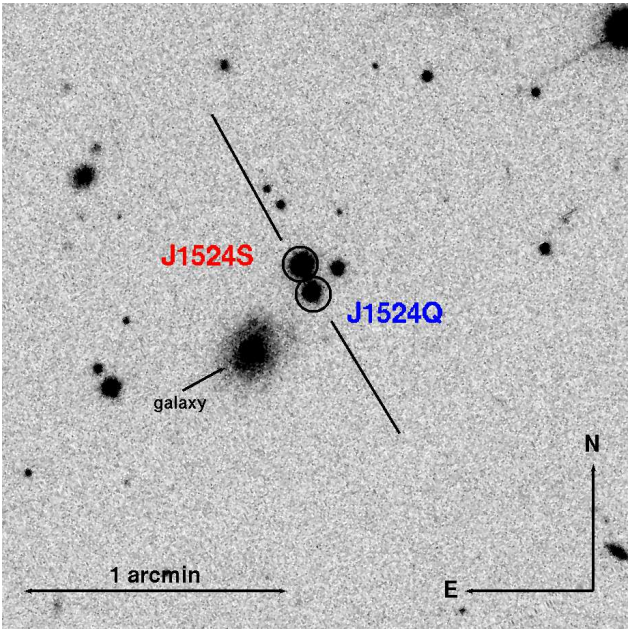}  
  \caption{The field of J1524QS as imaged in r--band by SDSS. The slit 
  orientations adopted in our observation is also plotted. The arrow
  points to the galaxy at ${\rm z}=0.0626$ located at $19\farcs0$ from
  J1524Q}
  \label{fig:field}
\end{figure}

In order to confirm the quasar association and to measure the relative 
systemic velocities from [OIII] lines (see \S\ref{sec:3.2}) we observed 
with the $1.82$\,m Asiago telescope the two quasars SDSS~J15244+3032 
(hereafter J1524Q) and RXS~J15244+3032 (hereafter J1524S), the nearest 
pair we have found (see Figure~\ref{fig:field}). 
In the VCV10 catalogue, it appears as a system of two bright radio quiet 
quasars separated in the sky by $\ad=6\farcs5$ with redshift ${\rm z}=0.274$ 
for J1524Q \citep{Adelman2006} and ${\rm z}=0.282$ for J1524S \citep{Zhao2000}. 
The two sources were detected in IR by the Two Micron All Sky Survey 
\citep[2MASS,][]{Skrutskie2006} with magnitudes: 
${\rm J}=15.28$, ${\rm H}=14.47$, and ${\rm K}=13.50$ for J1524Q, and
${\rm J}=14.18$, ${\rm H}=13.68$, and ${\rm K}=13.70$ for J1524S.
A source is present in the ROSAT All--Sky Survey Bright Source 
Catalogue \citep{Voges1999} at RA=15:24:28.6, DEC=+30:32:35.
Both J1524Q and J1524S are within the positional error of $12\arcsec$ 
($\sim2\arcsec$ from J1524Q and $\sim9\arcsec$ J1524S).

It is worth noting the presence of a galaxy at redshift \mbox{${\rm z}=0.0626$} 
\citep[RA=15:24:30,DEC=+30:32:24.2,][see Figure~\ref{fig:field}]{Aihara2011}
located at $19\farcs0$ from the quasar J1524Q. The corresponding projected 
separations of $23$\,kpc makes J1524Q a viable candidate to investigate the 
absorption features that the gaseous halos of the galaxy imprint on its 
spectra \citep[see e.g.,][]{Bahcall1969, Steidel1991, Chen2001, Chen2010, 
Adelberger2005, Farina2012}. Most prominent metal features are located in the 
Near UV, for instance the MgII doublet and the MgI line are shifted at 
2974\,\AA\ and at 3031\,\AA, respectively. Thus future observations with 
instruments like Hubble Space Telescope spectrographs STIS and COS, could 
increase the small number of metal absorption systems known at redshift 
${\rm z}<0.1$, where exquisite details of the galaxy population have been 
recorded \citep[see e.g.,][]{Kacprzak2011, Landoni2012}.

\section{Spectroscopic Data}\label{sec:data}

\begin{figure*}
  \centering
  \includegraphics[width=1.9\columnwidth]{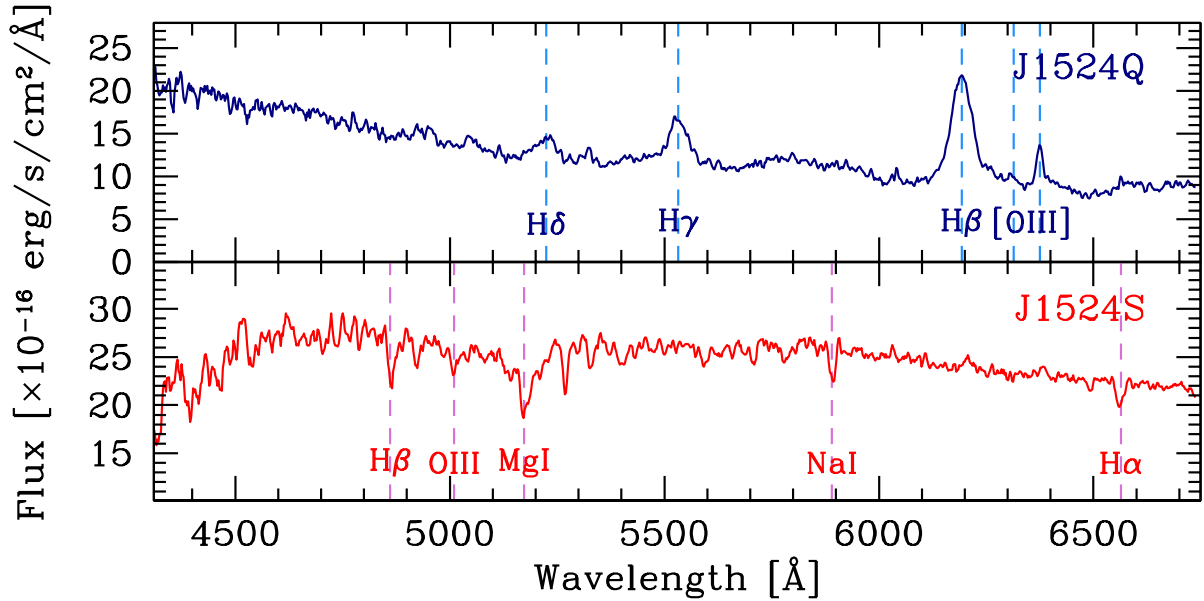}
  \caption{The spectra of J1524Q (Top Panel) and J1524S (Bottom Panel) 
  collected with AFOSC at the Asiago Observatory. 
  Most prominent absorption/emission lines are marked}
  \label{fig:spec}
\end{figure*}

J1524Q and J1524S were observed with the $1.82$\,m Asiago telescope located at 
the Cima Ekar Observatory on 20th April 2010. Data were gathered with the 
Asiago Faint Object Spectrograph and Camera (AFOSC) mounted in long--slit 
spectroscopy configuration with grism \#4 and $2\farcs1$ slit in the 
wavelength range from $\sim4300$\,\AA\ to $\sim6700$\,\AA. This yields a 
spectral resolution of ${\rm R}\sim300$. The targets are rather bright, 
thus the exposure time of 1800\,s (splitted into 3 exposures of 600\,s each)
with a seeing of $2\farcs5$ achieved an average signal--to--noise ratio per 
pixel of $\sim30$. The position angle of the slit was oriented so that the 
spectrum of the two objects could be acquired simultaneously (see 
Figure~\ref{fig:field}).

Standard \texttt{IRAF}\footnote{\texttt{IRAF} is distributed by the National 
Optical Astronomy Ob\-ser\-va\-to\-ries, which are operated by the Association 
of Universities for Research in Astronomy, Inc., under cooperative agreement 
with the National Science Foundation.} tools were adopted in the data 
reduction. Bias subtraction, flat field correction, image alignment and 
combination were performed with the \texttt{ccdred} package. Cosmic rays 
were cleaned by combining different exposures with the \texttt{crreject} 
algorithm. The \texttt{twodspec} and the \texttt{onedspec} packages were 
employed for the spectral extraction, the background subtraction and the 
calibrations both in wavelength and in flux. Residuals in wavelength 
calibrations are $\sim0.2$\,\AA. The spectra obtained are presented in 
Figure~\ref{fig:spec}.

\section{Reclassification of the pair}\label{sec:3.2}

\begin{table}
\centering
\caption{Properties of the detected emission lines
in the spectra of J1524Q.
Element (01);
peak wavelength of the line (02);
rest frame FWHM (03);
rest frame equivalent width of the line (04); and
estimated redshift (05)}\label{tab:lines}
\small
\begin{tabular}{lcccc}
\tableline
line                     & $\lambda_{\rm peak}$   & FWHM		   & EW 		     & z		\\
	                 & [\AA]		  & [km/s]		   & [\AA]		     &  		\\
(01)	                 & (02) 		  & (03)		   & (04)		     & (05)		\\
\tableline
$[$OIII$]_{\lambda5008}$ & 6378$\pm$1\phantom{.2} & \phantom{2}600$\pm$100 & \phantom{2}7$\pm$\phantom{2}1 & 0.2735	      \\
H$\beta$ (broad)	 & 6190$\pm$3\phantom{.2} & 3500$\pm$200	   & 62$\pm$12  		   & 0.273\phantom{0} \\
H$\gamma$ (broad)	 & 5530$\pm$2\phantom{.2} & 3200$\pm$250	   & 20$\pm$\phantom{2}5	   & 0.274\phantom{0} \\
H$\delta$ (broad)	 & 5226$\pm$3\phantom{.2} & 3500$\pm$400	   & \phantom{2}6$\pm$\phantom{2}2 & 0.274\phantom{0} \\
\tableline
\end{tabular}
\end{table}

The spectra of J1524S clearly shows a number of absorption features, the 
most noticeable of which are the rest--frame hydrogen and metal lines, 
indicating that it is likely a star of spectral type G (see 
Figure~\ref{fig:spec}). 
The erroneous classification of the source is confirmed by the study of 
\citet{Pickles2010}, who, from the analysis of the 2MASS and SDSS photometry, 
estimate that this object is a K0V star. We can argue that, although the 
finding chart published by \citep{Zhao2000} seems to point to J1524S, the 
authors have probably switched the two sources, and the X--ray emission 
observed by ROSAT was indeed associated with J1524Q. 
This is also supported by the lack of other sources within $20\arcsec$ from 
J1524S that have colour consistent with those of a quasar at 
${\rm z}\sim 0.28$.

It is well known that the different quasar emission lines can lead to 
redshift that differ by up to 1000\,km/s \citep[e.g.,][]{Tytler1992}. The 
most reliable estimate of the systemic redshift (${\rm z}_{\rm sys}$) comes 
from narrow forbidden lines, the most prominent of which are the [OIII] 
doublet at $\lambda=4949$\,\AA\ and $\lambda=5007$\,\AA\ 
\citep[e.g.,][]{Bonning2007, Hewett2010}. 
By measuring the position of the [OIII]$_{\lambda5008}$ line in our 
spectrum of J1524Q, we obtain ${\rm z}_{\rm sys}=0.2735\pm0.0003$ 
(see Table~\ref{tab:lines}). 
This agrees to 200\,km/s with the systemic redshift inferred from the 
SDSS spectrum by \citet[][${\rm z}_{\rm sys}=0.2743\pm0.0004$]{Hewett2010}.
The two redshift estimates are therefore marginally consistent within the 
uncertainties.
This value contrasts with the redshift presented by Zhao et al. 
(${\rm z}=0.282$). However, since the authors did not publish any 
indication on the uncertainty associated with their measure, we are 
not able to directly compare the two estimates. 

\section{Summary and Conclusions}\label{sec:conc}

In order to probe the physical association of the alleged low redshift 
quasar pair J1524QS present in the VCV10 catalogue, we have observed
the two sources with the Asiago telescope. The optical spectrum of J1524S 
shows a number of absorption features typical of a G spectral type star, 
ruling out the quasar classification proposed by \citet{Zhao2000} for this 
object.

It is worth noting that some other similar reclassification of alleged quasars 
in the V{\'e}ron-Cetty \& V{\'e}ron catalogues are present in the literature 
\citep[e.g.,][]{Decarli2009, Cupani2011}. Moreover \citet{Flesch2012} shows
that $\sim450$ quasars listed in the catalogue have erroneous astrometry
by more than $8\arcsec$ or are incorrect duplication of the same objects.
This suggests that, for the study of binary systems, one must take special 
care when considering the classification, the position, and the redshift 
of the sources listed in this catalogue,especially when no spectrum is 
publicly available.

\acknowledgments
We would like to thank the referee, A.~D.~Myers, for carefully reading
our manuscript.
For this work EPF was supported by Societ\`{a} Carlo Gavazzi S.p.A. and 
by Thales Alenia Space Italia S.p.A.
RD acknowledges funding from Germany's national research centre for 
aeronautics and space (DLR, project FKZ 50 OR 1104).
This research has made use of the NASA/IPAC Extragalactic Database (NED) 
which is operated by the Jet Propulsion Laboratory, California Institute 
of Technology, under contract with the National Aeronautics and Space 
Administration.
This research has made use of the VizieR catalogue access tool, CDS, 
Strasbourg, France.
Funding for SDSS-III has been provided by the Alfred P. Sloan Foundation, 
the Participating Institutions, the National Science Foundation, and the 
U.S. Department of Energy Office of Science. The SDSS-III web site is 
{\texttt http://www.sdss3.org/}.
SDSS-III is managed by the Astrophysical Research Consortium for the 
Participating Institutions of the SDSS-III Collaboration.

\end{document}